\documentclass{article}
\usepackage{spconf,amsmath,graphicx,hyperref}
\usepackage[numbers,sort&compress]{natbib}
\usepackage{afterpage}
\usepackage{booktabs}  % 用于三线表
\usepackage{array}      % 用于调整列宽
\usepackage{caption}    % 用于表格标题
\usepackage{graphicx}

\usepackage{amssymb}  % 提供 \checkmark
\usepackage{multirow} % 合并行
\usepackage{verbatim} % 提供 comment 环境

\usepackage{enumitem}
\usepackage{subcaption}

\title{MSF-SER: Enriching Acoustic Modeling with Multi-Granularity Semantics for Speech Emotion Recognition}
\name{
\begin{tabular}{@{}c@{}}
\itshape
Haoxun Li$^{1}$ \quad Yuqing Sun$^{1}$ \quad Hanlei Shi$^{1}$ \quad Yu Liu$^{1}$ \quad
Leyuan Qu$^{\star,1}$ \quad
Taihao Li$^{\star,1}$
\end{tabular}%
\thanks{$^{\star}$ Corresponding authors.}
\thanks{Haoxun Li, et al. Copyright 2026 IEEE. Personal use of this material is permitted. 
Permission from IEEE must be obtained for all other uses, including reprinting/republishing, 
creating new collective works, for resale or redistribution to servers or lists, or reuse of 
any copyrighted component of this work. DOI will be added upon IEEE Xplore publication.}
}
\address{$^{1}$ Hangzhou Institute for Advanced Study, University of Chinese Academy of Sciences}
\begin{document}
\ninept
%\fontsize{9.5pt}{11pt}\selectfont
%
\maketitle
\begin{abstract}
Continuous dimensional speech emotion recognition captures affective variation along valence, arousal, and dominance, providing finer-grained representations than categorical approaches. Yet most multimodal methods rely solely on global transcripts, leading to two limitations: (1) all words are treated equally, overlooking that emphasis on different parts of a sentence can shift emotional meaning; (2) only surface lexical content is represented, lacking higher-level interpretive cues. To overcome these issues, we propose MSF-SER (Multi-granularity Semantic Fusion for Speech Emotion Recognition), which augments acoustic features with three complementary levels of textual semantics—Local Emphasized Semantics (LES), Global Semantics (GS), and Extended Semantics (ES). These are integrated via an intra-modal gated fusion and a cross-modal FiLM-modulated lightweight Mixture-of-Experts (FM-MOE). Experiments on MSP-Podcast and IEMOCAP show that MSF-SER consistently improves dimensional prediction, demonstrating the effectiveness of enriched semantic fusion for SER.

\end{abstract}
\section{Introduction}
\label{sec:intro}

Speech Emotion Recognition (SER) has attracted increasing attention for its applications in human–computer interaction, mental health, and multimedia retrieval. Traditional approaches often rely on discrete emotion categories, which cannot fully capture the richness of human emotions. Recently, emotion dimensions have become popular and widely adopted, since they are more capable of capturing subtle variations and representing complex emotional states\cite{ref3}.

%cite ref7 delete in the cr 
Early dimensional SER systems largely relied on handcrafted acoustic and prosodic features combined with conventional machine learning models. Recent advances include the TF-Mamba architecture proposed by Zhao et al.\cite{ref9} for capturing temporal and frequency patterns, and efficient fine-tuning strategies for large SER models introduced by Aneesha et al\cite{ref10}.

Despite progress in speech-only approaches, semantic representations derived from acoustic signals remain limited, as they are easily affected by noise, speaker variability, and pronunciation differences\cite{ref11}. Text provides a more stable source of semantics, and pre-trained text models are trained on much larger corpora than speech models, yielding richer and more reliable representations\cite{text}. This motivates the incorporation of textual semantics to complement acoustic features. However, most existing multimodal studies still rely solely on global transcripts, which introduces two major issues.
% However, most existing multimodal studies only leverage global transcripts, which provide limited and shallow semantic cues.

% To move beyond this limitation, we argue that dimensional emotion recognition requires richer textual semantics. 
%In particular, global transcripts alone treat all words equally, overlooking that not every word contributes to emotion and that emphasis on different parts of a sentence can shift its affective meaning. 
(1) Global transcripts treat all words equally, overlooking the fact that not every word contributes to emotion and that emphasis on different parts of a sentence can shift its affective meaning. For example, in the sentence “I really didn’t mean that”, placing emphasis on “really” conveys a sense of sincerity or insistence, while stressing “didn’t” instead conveys denial or even defensiveness. Such differences can drastically alter the perceived emotion. This motivates the introduction of locally emphasized semantics, which are derived from the prosodically most prominent segments of speech. These emphasized regions often convey the speaker’s key semantic intent and exhibit stronger correlations with emotional expression. 

(2) Global transcripts only capture the explicit lexical content while lacking higher-level interpretive information, including but not limited to explanatory cues, situational factors, and paralinguistic attributes. Such information is essential for emotion recognition, since affective meaning is shaped not only by lexical content but also by inferred intent, communicative context, and speaker-specific characteristics. To address this gap, we incorporate prior-knowledge semantics, where external audio–language models provide extended and multi-dimensional interpretations of the audio, enhancing textual semantics with information inferred from audio. Along with local emphasized semantics and global contextual semantics, these prior-knowledge representations constitute our multi-granular semantic design.

To fully exploit enriched semantics, MSF-SER employs an intra-modal gated fusion to adaptively integrate different semantic levels, and a FiLM-modulated lightweight Mixture-of-Experts (FM-MOE) for cross-modal integration. Together, these components allow semantic cues to effectively guide acoustic representation learning, leading to consistent improvements in dimensional emotion recognition.

%To fully leverage these enriched semantics, we further introduce dedicated fusion strategies. An intra-modal gating mechanism adaptively integrates different levels of semantic features, while an inter-modal lightweight Mixture-of-Experts framework with FiLM modulation allows semantic cues to effectively guide acoustic representation learning. Together, these strategies ensure that the proposed semantic representations are maximally exploited, leading to substantial improvements in dimensional emotion recognition performance.

%\cite{ref21}

The main contributions of our work are summarized as follows:
%[noitemsep, topsep=0pt, leftmargin=*]
\begin{itemize}
\item[$\bullet$] For speech emotion recognition, we supplement acoustic modeling with richer semantic information by extracting features at three complementary granularities.
\item[$\bullet$] We employ FM-MOE for cross-modal fusion, while an intra-modal gating mechanism adaptively integrates global and local semantics to enhance acoustic feature learning.
\item[$\bullet$] MSF-SER achieves top-performing results across MSP-Podcast and IEMOCAP, thereby demonstrating the effectiveness and generalizability of our method.

%Our approach, MSF-SER, outperforms the top-performing method\cite{ref22} from SERNC challenge\cite{ref14} on the MSP-Podcast dataset and demonstrates effectiveness on the IEMOCAP dataset.
\end{itemize}

\begin{figure*}[t!]
    \centering
    \includegraphics[width=0.9\textwidth]{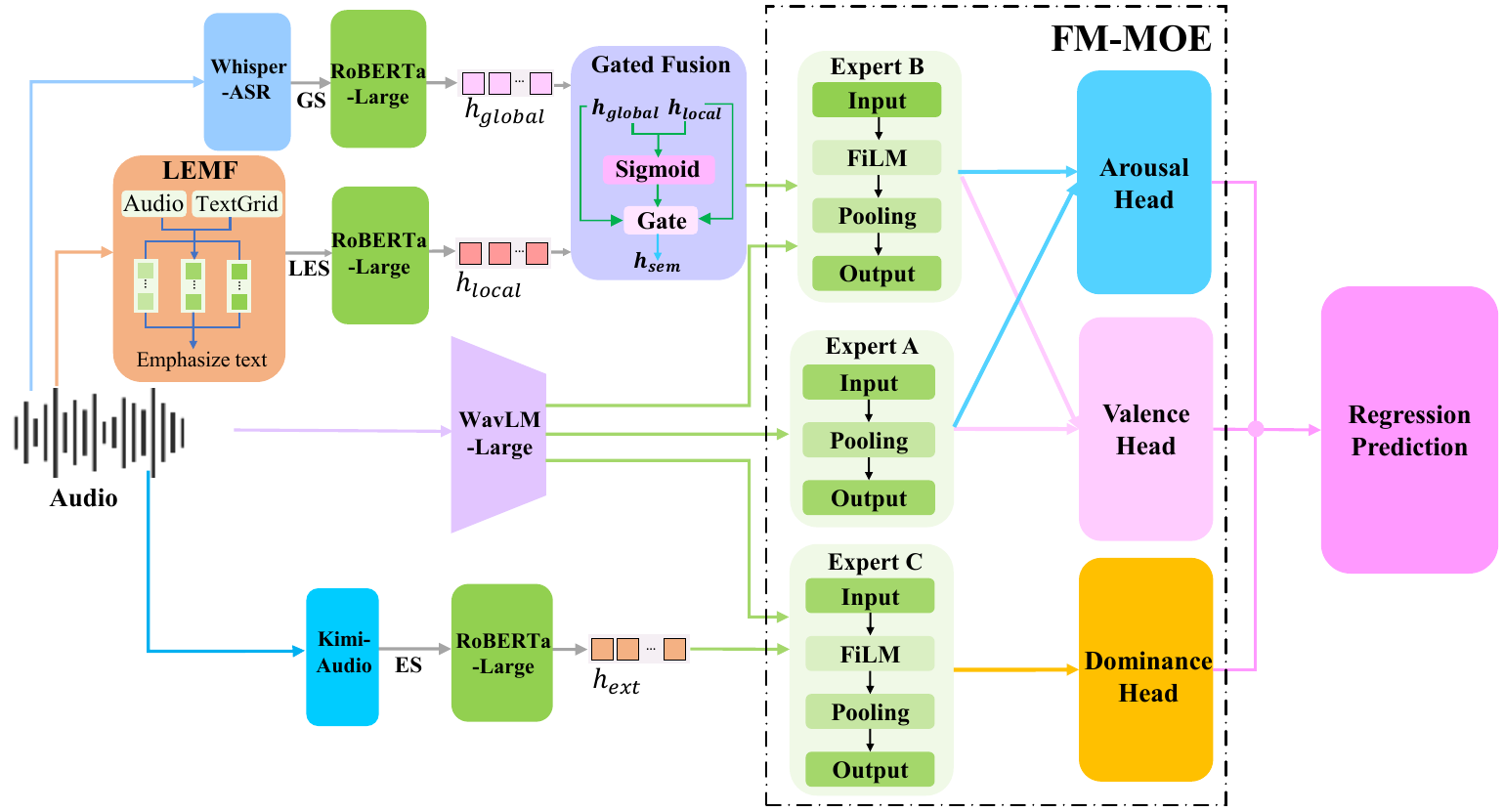}
    \vspace{-0.5em}
    \caption{Our proposed framework adopts acoustic features as the backbone, while three levels of semantic features are incrementally incorporated as auxiliary signals. An intra-modal gating mechanism fuses global and local semantics, and cross-modal integration between acoustic and semantic representations is achieved through FM-MOE.}
    \label{fig:overview}
\end{figure*}

\section{Method}
\label{sec:format}

\subsection{Overview}
\label{ssec:subhead}

% Our model builds upon the Interspeech 2025 baseline system\cite{ref14}, which employs a fine-tuned WavLM-large architecture\cite{ref23}. In the baseline, raw audio is encoded into frame-level acoustic representations by the WavLM encoder, followed by attentive statistics pooling\cite{ref24} to compute weighted mean and standard deviation, and a fully connected layer for predicting valence, arousal, and dominance.

Our baseline employs a fine-tuned WavLM-large architecture, where raw audio is first encoded into frame-level acoustic representations by the WavLM encoder. The outputs are then aggregated using attentive statistics pooling to compute weighted mean and standard deviation, followed by a fully connected layer that predicts valence, arousal, and dominance.

On this foundation, we introduce multi-granularity semantic features: LES, GS, and ES, as illustrated in Fig.~\ref{fig:overview}. These features interact with acoustic features through a two-stage fusion framework. For intra-modal fusion, we design a gated mechanism to adaptively integrate LES and GS, enhancing model sensitivity and stability. For inter-modal fusion, we propose FM-MOE. Specifically, Feature-wise Linear Modulation (FiLM) applies dimension-wise scaling and shifting to acoustic representations conditioned on semantic features, ensuring fine-grained cross-modal interaction. Within FM-MOE, FiLM-modulated experts specialize in different emotional dimensions, and their outputs are adaptively combined through different routing weights. This design enables semantic cues to effectively guide acoustic learning while capturing dimension-specific dependencies.

%On this foundation, we introduce multi-granularity semantic features: LES, GS, and ES, as illustrated in Fig \ref{fig:overview}. These features interact with acoustic features through a two-stage fusion framework. For intra-modal fusion, we design a gated mechanism to adaptively integrate LES and GS, enhancing model sensitivity and stability. For inter-modal fusion, we employ Feature-wise Linear Modulation layers\cite{film} together with a lightweight Mixture-of-Experts\cite{ref21}, enabling semantic cues to effectively guide acoustic representation learning and capture dimension-specific dependencies.

%On this foundation, we introduce multi-granularity semantic features: Local Emphasized Semantics (LES), Global Semantics (GS), and Extended Semantics (ES). These features interact with acoustic features through a three-level fusion mechanism. For intra-modal fusion, we design a gated fusion mechanism to enhance model sensitivity and discriminative power while maintaining training stability. For cross-modal interaction, we employ feature-wise Linear Modulation layers\cite{film} to improve multimodal fusion efficiency. To account for dimension-specific dependencies, a lightweight Mixture-of-Experts \cite{ref21} mechanism is further applied.

%The overall framework is illustrated in Fig \ref{fig:overview}. 

\subsection{Multi-Granularity Semantic Feature Extraction}
\label{ssec:subhead}

%In spontaneous speech, emphasized regions typically convey critical semantic information.

\textbf{Local Emphasized Semantics (LES)}. Prior work\cite{seshadri2021emphasis} indicates that prosodic prominence mainly arises from pitch, energy, and duration. However, in noisy conditions, emphasis detection models trained on clean data, such as EmphaClass\cite{EmphaClass}, often degrade. To make emphasis modeling more robust, we design a \textbf{Local Emphasis Modeling Framework (LEMF)}, which identifies prosodically prominent words and transforms them into LES. LES highlight the most salient parts of speech that often carry the speaker’s key intent and emotional salience. The specific implementation of LEMF is as follows:

Given a sentence $u=\{w_1, \dots, w_N\}$ with audio signal $x(t)$, where $t$ denotes time, and its MFA-based TextGrid alignment\cite{ref26}, we extract three prosodic features for each word $w$:

\begin{itemize}[noitemsep, topsep=0pt, leftmargin=*]
    \item \textbf{Pitch} ($f_{\text{pitch}}(w)$): log-F0 via the PyWorld Harvest algorithm;  
    \item \textbf{Energy} ($f_{\text{energy}}(w)$): L2 norm of Short-Time Fourier Transform (STFT) using a 20ms window;  
    \item \textbf{Duration} ($f_{\text{duration}}(w)$): mean phoneme duration in $w$.  
\end{itemize}

To ensure comparability of features within a sentence, we compute sentence-level statistics (mean $\mu_i$, standard deviation $\sigma_i$) for pitch $f_{\text{pitch}}(w)$, energy $f_{\text{energy}}(w)$ and duration $f_{\text{duration}}(w)$. Each word-level feature is then normalized using Z-score standardization, as shown in Equation \ref{eq:z_normalization}.

\vspace{-1em}
\begin{equation}
\label{eq:z_normalization}
z_i(w) = \frac{f_i(w) - \mu_i}{\sigma_i}, \quad i \in \{\text{pitch}, \text{energy}, \text{duration}\}
\end{equation}

where $f_i(w)$ denotes the raw feature of $w$ in the $i$-th prosodic dimension. Specifically, pitch corresponds to the maximum frame-level F0, energy to the mean frame energy, and duration to average phoneme length.

Subsequently, we perform a weighted fusion of the three prosodic features to obtain the emphasis score s(w) for each word $w$ as shown in Equation \ref{2}:
\begin{equation}
\label{2}
s(w) = \alpha \cdot z_{\text{pitch}}(w) + \beta \cdot z_{\text{energy}}(w) + \gamma \cdot z_{\text{duration}}(w)
\end{equation}
where $(\alpha, \beta, \gamma) = (1.0, 1.2, 0.8)$. The word with the highest score, along with its two adjacent words, forms the emphasis segment. This segment is then encoded by RoBERTa-Large\cite{ref28} to derive LES features.  

\vspace{0.5em}
\noindent
\textbf{Global Semantics (GS)}. We derive sentence-level semantics by transcribing speech with Whisper-ASR\cite{ref29} and encoding the text using RoBERTa-Large.  

\vspace{0.5em}
\noindent
\textbf{Extended Semantics (ES)}. Leveraging recent advances in large-scale audio understanding, we employ Kimi-Audio\cite{ref20} to generate six categories of extended speech and emotion-related information, as shown in Table~\ref{tab:extended_info}. These outputs are concatenated into descriptive text and encoded with RoBERTa-Large, yielding enriched semantic representations. For example, an extended semantic description may state: “This is a male speaker, expressing frustrated (categorized as Angry), in a confrontation or argument in a public setting. The speaker is expressing anger due to a perceived lack of understanding or compliance from the other party. The speech is characterized by rising intonation, assertive tone, and a sense of urgency.”

\begin{table}[htbp]
\centering
\renewcommand{\arraystretch}{1} % 调整行距
\caption{Extended semantics information categories generated by Kimi-Audio.}
\vspace{-0.5em}
\label{tab:extended_info}
\small  % 或 \footnotesize / \scriptsize
\begin{tabular}{p{3.5cm}p{4.2cm}}
\toprule
\textbf{Category} & \textbf{Description} \\
\midrule
Free emotion label & Open-domain emotion prediction \\
Constrained emotion label & Mapping to dataset labels \\
Emotion explanation & Reason for emotion prediction \\
Scenario & Potential situational context \\
Paralinguistics & Paralinguistic information \\
Gender & Gender information \\
\bottomrule
\end{tabular}
\end{table}

\vspace{-1.7em}
\subsection{Intra-modal Gated Fusion}
\label{ssec:subhead}

Locally emphasized semantics capture salient emotional cues but lack holistic coverage, whereas global semantics provide comprehensive context but may introduce redundancy and noise. To integrate their complementary strengths, we employ a gated fusion mechanism, as shown in Equation \ref{3}.

\vspace{-1.7em}
\begin{align}
\label{3}
h_{\text{sem}} &= g \cdot h_{\text{local}} + (1 - g) \cdot h_{\text{global}} \\
g &= \sigma(W [h_{\text{local}}; h_{\text{global}}] + b)
\end{align}
where $h_{\text{sem}}$ is the fused representation, $h_{\text{local}}$ and $h_{\text{global}}$ denote local and global semantics, respectively, $g$ is the learned gating weight, $\sigma$ is the Sigmoid activation, and $W$, $b$ are trainable parameters.

\vspace{-0.5em}

%\subsection{Modal FiLM Modulation Layer Integration}
\subsection{FM-MOE}
%\subsection{FiLM-Modulated Lightweight Mixture-of-Experts (FM-MOE)}
\label{ssec:subhead}

In dimensional emotion recognition, acoustic features remain the primary source of information, while text semantics provide complementary cues. To facilitate fine-grained cross-modal interaction, we employ a Feature-wise Linear Modulation (FiLM) layer\cite{film}, which applies dimension-wise scaling and shifting to acoustic representations conditioned on semantic features, as shown in Equation \ref{5}. This design prioritizes speech as the dominant modality while leveraging semantics as auxiliary guidance, thereby mitigating the impact of semantic noise and improving fusion effectiveness.
%In dimensional emotion recognition, acoustic features dominate while text semantics serve as complementary cues. To enable fine-grained cross-modal interaction, we adopt a Feature-wise Linear Modulation layer\cite{film}, which applies dimension-wise scaling and shifting to acoustic features conditioned on semantics, as shown in Equation \ref{5}. This design follows a ``speech-primary, semantics-secondary'' paradigm, mitigating semantic noise while enhancing fusion efficiency.  

\vspace{-0.4em}
\begin{equation}
\label{5}
\tilde{h}_{\text{audio}} = \gamma \odot h_{\text{audio}} + \beta
\end{equation}

\vspace{-1.3em}

\begin{equation}
\gamma, \beta = \text{MLP}(h_{\text{sem}})
\end{equation}
where $\tilde{h}_{\text{audio}}$ denotes the semantically modulated acoustic feature, $h_{\text{audio}}$ represents the input acoustic feature, $\gamma$ and $\beta$ are the scaling and bias parameters learned through the Multi-Layer Perceptron (MLP), and $h_{\text{sem}}$ is the semantic feature vector.

%\subsection{Lightweight Expert Mixture}
%\label{ssec:subhead}

To capture the distinct cross-modal dependencies of different emotional attributes, we propose a lightweight Mixture-of-Experts module following the shared WavLM backbone. It consists of three experts: an \textit{acoustic-only} expert (Expert A), a \textit{textual semantic} expert (Expert B) incorporating fused local and global semantics, and an \textit{extended semantic} expert (Expert C) that integrates additional prior knowledge.
%It consists of three experts: an \textit{acoustic-only} expert (Expert A), a \textit{global semantic} expert (Expert B) incorporating utterance-level semantics, and an \textit{extended semantic} expert (Expert C) that integrates additional prior knowledge.

Subsequently, for each emotional dimension $d \in \{\text{V}, \text{A}, \text{D}\}$, the output is a weighted combination of the experts, as shown in Equation \ref{7}. 

\vspace{-1em}

\begin{equation}
\label{7}
\hat{y}^{(d)} = \sum_{k=1}^{3} \pi_k^{(d)} \cdot f_k(h)
\end{equation}
where $f_k$ denotes the $k$-th expert and $\pi_k^{(d)}$ represents the learnable routing weight. The routing weights allow different emotional dimensions to adaptively emphasize relevant experts (e.g., Valence and Arousal typically rely more on Experts A and B, while Dominance emphasizes Experts C).
%Valence and Arousal are jointly modeled by Expert A and Expert B, while Dominance is modeled by Expert A and Expert C.
\vspace{-0.5em}

\begin{table*}[!t]
\centering
\small
\caption{Ablation results on the MSP-Podcast development set. Results for [17] are cited from the IS25-SER Challenge test set and provided solely as a SOTA reference (not directly comparable).}
\begin{tabular}{@{}ccc|cc|cccc|cccc@{}}
\toprule
\multicolumn{3}{c|}{\textbf{Semantic}} & 
\multicolumn{2}{c|}{\textbf{Intra-Modal Fusion}} & 
\multicolumn{4}{c|}{\textbf{Inter-Modal Fusion}} & 
\multicolumn{4}{c}{\textbf{Evaluation Metrics}} \\
\cmidrule(lr){1-3}\cmidrule(lr){4-5}\cmidrule(lr){6-9}\cmidrule(lr){10-13}
GS & LES & ES & Attention & Gated & Concat & Attention & FiLM & FM-MOE & $\text{CCC}_V$ & $\text{CCC}_A$ & $\text{CCC}_D$ & $\text{CCC}_{avg}$ \\
\midrule\midrule
 &  &  &  &  &  &  &  &  & 0.725 & 0.660 & 0.592 & 0.659 \\
\checkmark &  &  &  &  & \checkmark &  &  &  & 0.710 & 0.658 & 0.591 & 0.653 \\
\checkmark &  &  &  &  &  & \checkmark &  &  & 0.697 & 0.655 & 0.587 & 0.646 \\
\checkmark &  &  &  &  &  &  & \checkmark &  & 0.741 & 0.668 & 0.608 & 0.672 \\
 & \checkmark &  &  &  &  &  & \checkmark &  & 0.739 & 0.665 & 0.610 & 0.671 \\
 &  & \checkmark &  &  &  &  & \checkmark &  & 0.728 & 0.652 & 0.630 & 0.670 \\
\checkmark & \checkmark &  & \checkmark &  &  &  & \checkmark &  & 0.745 & 0.670 & 0.606 & 0.677 \\
\checkmark & \checkmark &  &  & \checkmark &  &  & \checkmark &  & 0.756 & 0.675 & 0.612 & 0.681 \\
\checkmark & \checkmark & \checkmark &  & \checkmark &  &  & \checkmark &  & 0.750 & 0.670 & 0.622 & 0.681 \\
\midrule

\multicolumn{9}{l|}{ML-Adapt\cite{goncalves2024bridging}} & 0.647 & 0.680 & 0.616 & 0.648 \\
\multicolumn{9}{l|}{LoRA-SER\cite{chochlakis2025modality} } & 0.701 & 0.682 & 0.616 & 0.666 \\
\multicolumn{9}{l|}{DEER\cite{wu2023estimating}} & 0.688 & \textbf{0.696} & 0.612 & 0.665 \\
\multicolumn{9}{l|}
{PCM-le-noNorm\cite{ryu2025pitch}} & 0.658& 0.675 & 0.626 & 0.653 \\
\multicolumn{9}{l|}{SERNC Top-Performing Model\cite{ref13} (test-set only)} & 0.758 & 0.683 & 0.615 & 0.685 \\
\midrule
\checkmark & \checkmark & \checkmark &  & \checkmark &  &  &  & \checkmark & \textbf{0.759} & 0.685 & \textbf{0.631} & \textbf{0.692} \\
\bottomrule
\end{tabular}
\label{tab:experiments}
\end{table*}

\section{Experiments}
\label{sec:typestyle}

\subsection{Experimental Setup}

We conducted experiments on MSP-Podcast v1.12\cite{busso2025msp} and IEMO-CAP\cite{busso2008iemocap}. MSP-Podcast v1.12 is a large-scale corpus of spontaneous podcast speech with 84,260 training and 31,961 development utterances, annotated with ten categorical labels and dimensional ratings (arousal, valence, dominance, 1–7 scale), and officially split into training, development, and three test sets. IEMOCAP contains 10,039 utterances from ten actors across five sessions of scripted and spontaneous dialogues, annotated with categorical emotions and dimensional ratings on a 1–5 scale.

We train with AdamW ($1\times10^{-5}$ learning rate), batch size 32, and gradient accumulation of 4. We optimize emotional attribute regression by minimizing the Concordance Correlation Coefficient (CCC) loss\cite{lawrence1989concordance}. Each emotional dimension is predicted by an independent two-layer regression head with dropout 0.5 and layer normalization. For features, we use WavLM-Large as the acoustic encoder (hidden size 1024), freezing the CNN front-end and fine-tuning the Transformer layers, and RoBERTa-Large (1024 hidden size) for textual semantics. Training is performed on 8 RTX 4090 GPUs.
%We conducted experiments on MSP-Podcast v1.12 and IEMOCAP. MSP-Podcast v1.12 is a large-scale corpus of natural, spontaneous emotional speech collected from podcast recordings, containing 84,260 training and 31,961 development utterances. Each utterance is annotated with ten categorical labels and dimensional ratings of arousal, valence, and dominance on a 1–7 scale. The corpus is officially split into training, development, and three test sets. In contrast, IEMOCAP consists of 10,039 utterances from ten professional actors, with each pair forming one session that includes both scripted and spontaneous dialogues. Each utterance is annotated with categorical emotion labels as well as dimensional ratings of valence, arousal, and dominance on a 1–5 scale.

%We adopt the AdamW optimizer with a learning rate of $1\times10^{-5}$, a batch size of 32, and gradient accumulation over 4 steps. Each emotional dimension (arousal, valence, dominance) is assigned an independent regression head, implemented as a two-layer fully connected network with dropout=0.5 and layer normalization, enabling dimension-specific prediction strategies. For feature extraction, we use WavLM-Large as the acoustic encoder (hidden size = 1024), where the CNN-based feature encoder is frozen and only the Transformer-based contextual encoder is fine-tuned. For textual semantics, we employ RoBERTa-Large with a 1024-dimensional hidden size. Training is conducted on 8 RTX 4090 GPUs.

\subsection{Performance on Emotional Attribute Prediction}

Table \ref{tab:experiments} presents our results on the MSP-Podcast development set, alongside a test-set reference [17] for context. Model performance is measured using CCC scores\cite{lawrence1989concordance}.
%Our evaluation results on the MSP-Podcast dataset are reported on the development set, as shown in Table \ref{tab:experiments}. Model performance is measured using CCC scores\cite{lawrence1989concordance}.
%which jointly accounts for both correlation and mean squared difference between predictions and ground truth, making it the standard metric for dimensional emotion recognition.

We adopt a baseline architecture built on WavLM-Large as the speech encoder, followed by an attentive statistics pooling layer and fully connected layers to perform regression prediction. The results demonstrate that incorporating semantic information at three different granularities consistently improves recognition performance. Specifically, introducing GS and LES yields greater gains for arousal and valence prediction, while ES mainly enhances dominance regression. Regarding inter-modal fusion strategies, both concatenation and attention degrade performance. We attribute this to the fact that, in speech-dominant tasks, these methods allow noisy textual semantics to interfere with the intrinsic structure of acoustic representations. In contrast, FiLM-based fusion effectively mitigates such noise and leads to more robust predictions.
Moreover, applying intra-modal fusion on GS and LES further improves results compared to using either individually, with the gated approach outperforming the attention-based method. Finally, when all three levels of semantic information are jointly introduced and integrated with FM-MOE, our model achieves the best overall performance.
%surpassing the SERNC top-performing method.
%\vspace{-4pt}
%\begin{comment}
% ---- 客观评价表 (Objective Evaluation) ----
\begin{table}[ht]
\vspace{-4pt}
\caption{Comparison of different models on the IEMOCAP dataset.}
\label{tab:objective_evaluation}
\centering

\resizebox{0.46\textwidth}{!}{  % Resize to fit within the page width
\fontsize{10pt}{10pt}\selectfont  % 调整字体大小
\begin{tabular}{ l c c c c c c }
\toprule
\textbf{Model} & \textbf{$\text{CCC}_{V}$} & \textbf{$\text{CCC}_{A}$} & \textbf{$\text{CCC}_{D}$} & \textbf{$\text{CCC}_{avg}$} \\
\midrule

Baseline \cite{ref14}       & 0.552 & 0.678  & 0.583 &0.604  \\
KNN-VC\cite{mote2024unsupervised}& 0.568 & 0.656  & 0.485 &0.570  \\
WavLM-LR\cite{naini2024generalization}& 0.625 & 0.675  & 0.599 &0.633  \\
DEER \cite{wu2023estimating}        & 0.625 & 0.711 & 0.548 & 0.628  \\
PCM-le-noNorm\cite{ryu2025pitch} & 0.630 & \textbf{0.717} & 0.555 & 0.634  \\
\midrule
\textbf{MSF-SER}           &\textbf {0.632} & 0.680 & \textbf{0.601} & \textbf{0.638}  \\
\bottomrule
\end{tabular}
}
\end{table}
%\vspace{-0.5pt}

\begin{figure}[htbp] 
    \vspace{-1em} % 上移图片
    \centering
    \begin{subfigure}{1\linewidth}
        \centering
        \includegraphics[width=\linewidth]{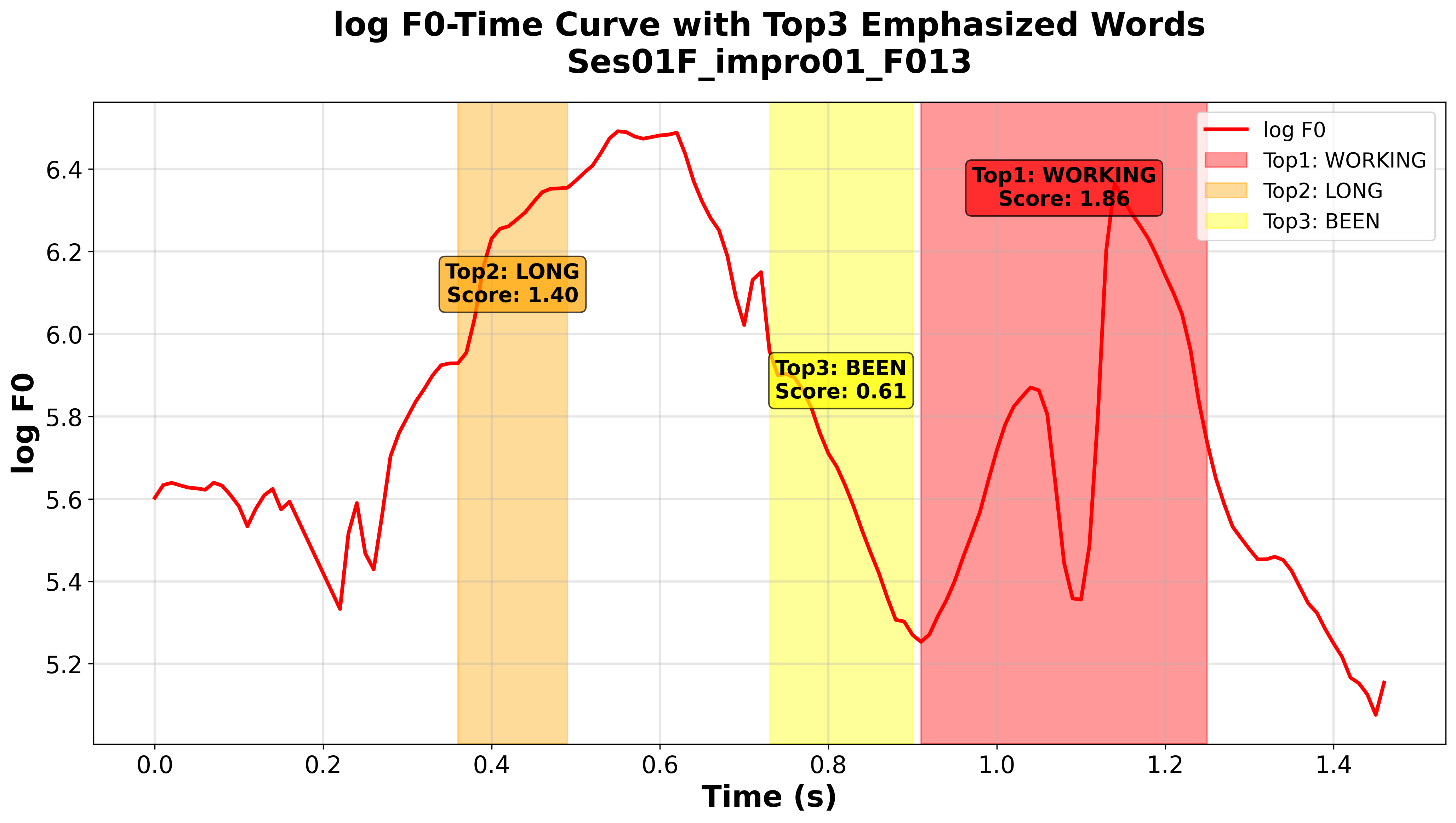}
    \end{subfigure}
    
    \vspace{0.5em}
    \begin{subfigure}{1\linewidth}
        \centering
        \includegraphics[width=\linewidth]{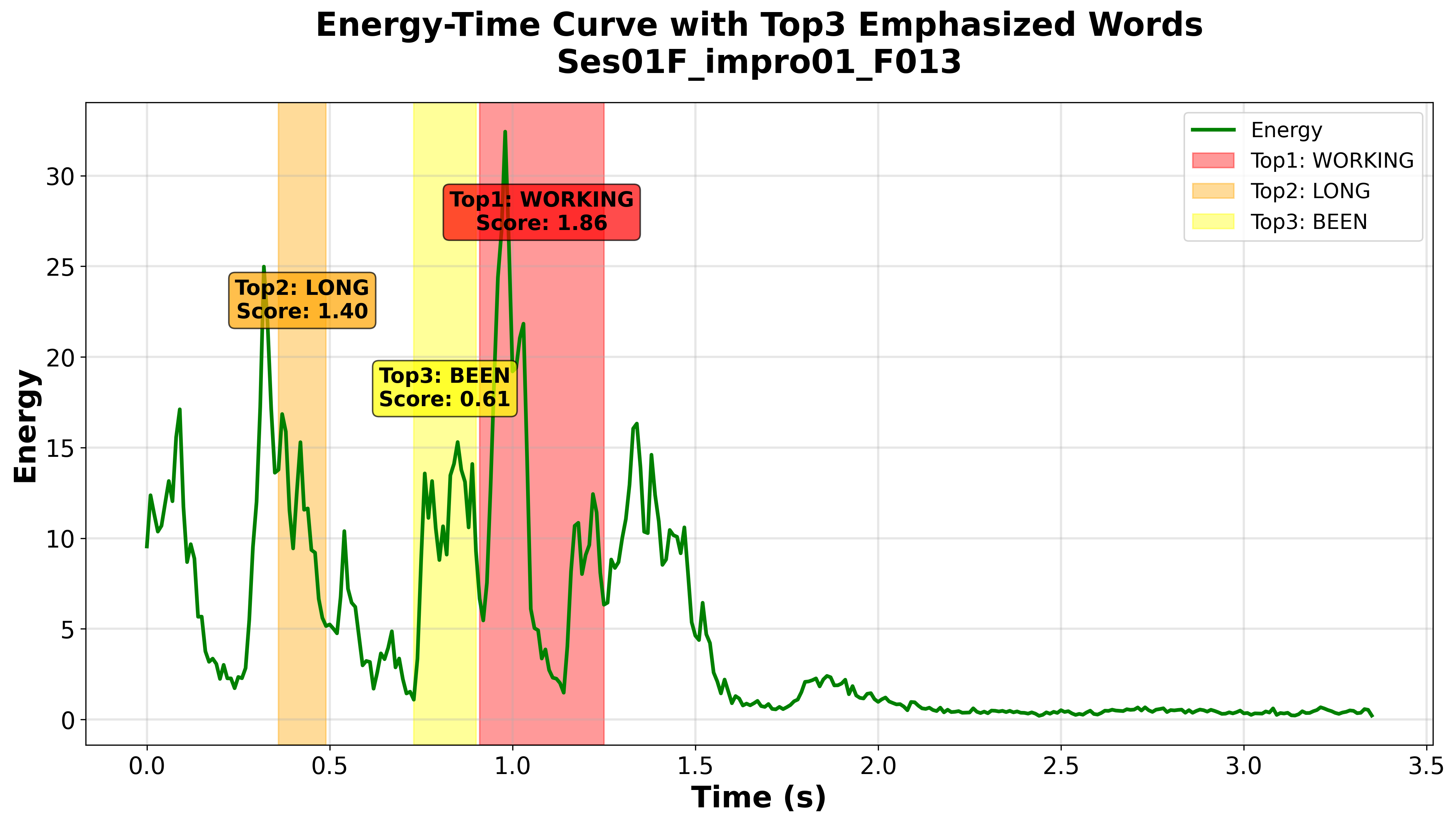}
    \end{subfigure}
    
    \caption{Illustration of emphasis detection by LEMF. The top figure shows the log-F0 curve over time with the top-3 emphasized words highlighted, while the bottom figure shows the corresponding energy-time curve.}
    \label{fig:prosody_features}
    \vspace{-0.5em} % 减少图片底部空白
\end{figure}

We further evaluated our model on the IEMOCAP dataset to validate its effectiveness. We adopt five-fold cross-validation, using four sessions for training and the remaining session for testing in each fold, ensuring strict speaker independence. As shown in Table 3, we compared our approach with multiple models. MSF-SER achieves the highest CCC scores on valence, dominance and the overall average, demonstrating its advantage in capturing subtle emotional cues across dimensions.

\subsection{LEMF}

LEMF is designed to improve emphasis detection in spontaneous speech, where noise and speaking style variations often degrade performance. By leveraging prosodic features such as pitch, energy, and duration, LEMF can more accurately capture emphasized regions, yielding higher accuracy and robustness. As illustrated in Fig.~\ref{fig:prosody_features} ,this feature-based approach not only enhances adaptability in spontaneous conditions but also provides reliable local semantic cues for downstream emotion recognition tasks.

% \begin{figure}[t!] % 使用 t! 强制顶部放置
%     \vspace{-1em} % 上移图片
%     \centering
%     \begin{subfigure}{1\linewidth}
%         \centering
%         \includegraphics[width=\linewidth]{Ses01F_impro01_F013_log_F0_time.png}
%     \end{subfigure}
    
%     \vspace{0.5em}
%     \begin{subfigure}{1\linewidth}
%         \centering
%         \includegraphics[width=\linewidth]{Ses01F_impro01_F013_energy_time.png}
%     \end{subfigure}
    
%     \caption{Illustration of emphasis detection by LEMF. The top figure shows the log-F0 curve over time with the top-3 emphasized words highlighted, while the bottom figure shows the corresponding energy-time curve.}
%     \label{fig:prosody_features}
%     \vspace{-0.5em} % 减少图片底部空白
% \end{figure}

\section{Conclusion}

In this paper, we proposed MSF-SER, a multi-granularity semantic fusion framework for speech emotion recognition. By introducing LES, GS, and ES, and integrating them through gated fusion and FM-MOE, our approach enhances acoustic modeling with enriched textual cues. Experiments on MSP-Podcast and IEMOCAP show consistent improvements over strong baselines, establishing MSF-SER as a top-performing system. In future work, we plan to extend this design to cross-lingual scenarios and explore its integration with additional modalities such as visual signals.

\label{sec:majhead}

\begin{comment}

\end{comment}

\section*{Acknowledgments}
This work was supported in part by the Scientific Research Starting Foundation of Hangzhou Institute for Advanced Study (2024HIASC2001), in part by the National Natural Science Foundation of China (No. 62506091), in part by the Zhejiang Provincial Natural Science Foundation of China (No.\ LQN25F020001), and in part by the Key R\&D Program of Zhejiang (2025C01104).
% The authors declare no competing interests.

\noindent\textbf{Use of Generative AI and AI-Assisted Tools.}
Language editing in throughout the manuscript was assisted by ChatGPT (OpenAI) to improve grammar and clarity; all scientific content was authored by the authors.
During implementation, the authors used Cursor (an AI code assistant) for debugging support; no AI-generated code, figures, tables, or text were included in the manuscript.
All AI-assisted outputs were reviewed and verified by the authors, who take full responsibility for the content.

\section*{Compliance with Ethical Standards}
This study involved no human or animal subjects and did not require ethics approval.

\bibliographystyle{IEEEbib}
\bibliography{strings,refs}

@article{ref3,
  title={Survey on speech emotion recognition: Features, classification schemes, and databases},
  author={El Ayadi, Moataz and Kamel, Mohamed S and Karray, Fakhri},
  journal={Pattern recognition},
  volume={44},
  number={3},
  pages={572--587},
  year={2011},
  publisher={Elsevier}
}

@inproceedings{ref9,
  title={Temporal-frequency state space duality: An efficient paradigm for speech emotion recognition},
  author={Zhao, Jiaqi and Wang, Fei and Li, Kun and others},
  booktitle={ICASSP 2025-2025 IEEE International Conference on Acoustics, Speech and Signal Processing (ICASSP)},
  pages={1--5},
  year={2025},
  organization={IEEE}
}

@inproceedings{ref10,
  title={Efficient Finetuning for Dimensional Speech Emotion Recognition in the Age of Transformers},
  author={Sampath, Aneesha and Tavernor, James and Provost, Emily Mower},
  booktitle={ICASSP 2025-2025 IEEE International Conference on Acoustics, Speech and Signal Processing (ICASSP)},
  pages={1--5},
  year={2025},
  organization={IEEE}
}

@article{ref11,
  title={Learning fine-grained cross modality excitement for speech emotion recognition},
  author={Li, Hang and Ding, Wenbiao and Wu, Zhongqin and others},
  journal={arXiv preprint arXiv:2010.12733},
  year={2020}
}

@article{ref13,
  title={Developing a High-performance Framework for Speech Emotion Recognition in Naturalistic Conditions Challenge for Emotional Attribute Prediction},
  author={Lertpetchpun, Thanathai and Feng, Tiantian and Byrd, Dani and others},
  journal={arXiv preprint arXiv:2506.10930},
  year={2025}
}

@inproceedings{ref14,
  title={The interspeech 2025 challenge on speech emotion recognition in naturalistic conditions},
  author={Naini, A Reddy and Goncalves, Lucas and Salman, Ali N and others},
  year={2025},
  organization={Interspeech}
}

@article{ref20,
  title={Kimi-audio technical report},
  author={Ding, Ding and Ju, Zeqian and Leng, Yichong and others},
  journal={arXiv preprint arXiv:2504.18425},
  year={2025}
}

@inproceedings{film,
  title={Film: Visual reasoning with a general conditioning layer},
  author={Perez, Ethan and Strub, Florian and De Vries, Harm and others},
  booktitle={Proceedings of the AAAI conference on artificial intelligence},
  volume={32},
  number={1},
  year={2018}
}

@article{ref26,
  title={Multi-matrix factorization attention},
  author={Hu, Jingcheng and Li, Houyi and Zhang, Yinmin and others},
  journal={arXiv preprint arXiv:2412.19255},
  year={2024}
}

@article{ref28,
  title={Roberta: A robustly optimized bert pretraining approach},
  author={Liu, Yinhan and Ott, Myle and Goyal, Naman and others},
  journal={arXiv preprint arXiv:1907.11692},
  year={2019}
}

@inproceedings{ref29,
  title={Robust speech recognition via large-scale weak supervision},
  author={Radford, Alec and Kim, Jong Wook and Xu, Tao and others},
  booktitle={International conference on machine learning},
  pages={28492--28518},
  year={2023},
  organization={PMLR}
}

@inproceedings{chochlakis2025modality,
  title={Modality-Agnostic Multimodal Emotion Recognition using a Contrastive Masked Autoencoder},
  author={Chochlakis, Georgios and Iqbal, Turab and Kang, Woo Hyun and others},
  booktitle={Proc. Interspeech 2025},
  pages={3005--3009},
  year={2025}
}

@inproceedings{goncalves2024bridging,
  title={Bridging emotions across languages: Low rank adaptation for multilingual speech emotion recognition},
  author={Goncalves, Lucas and Robinson, Donita and Richerson, Elizabeth and others},
  booktitle={Proc. Interspeech 2024},
  pages={4688--4692},
  year={2024}
}

@inproceedings{ryu2025pitch,
  title={Pitch Contour Model (PCM) with Transformer Cross-Attention for Speech Emotion Recognition},
  author={Ryu, Minji and Hur, Ji-Hyeon and Kim, Sung Heuk and others},
  booktitle={Proc. Interspeech 2025},
  pages={4353--4357},
  year={2025}
}

@article{wu2023estimating,
  title={Estimating the uncertainty in emotion attributes using deep evidential regression},
  author={Wu, Wen and Zhang, Chao and Woodland, Philip C},
  journal={arXiv preprint arXiv:2306.06760},
  year={2023}
}

@inproceedings{mote2024unsupervised,
  title={Unsupervised domain adaptation for speech emotion recognition using K-Nearest neighbors voice conversion},
  author={Mote, Pravin and Sisman, Berrak and Busso, Carlos},
  booktitle={Proceedings of INTERSPEECH},
  year={2024}
}

@inproceedings{naini2024generalization,
  title={Generalization of self-supervised learning-based representations for cross-domain speech emotion recognition},
  author={Naini, Abinay Reddy and Kohler, Mary A and Richerson, Elizabeth and others},
  booktitle={ICASSP 2024-2024 IEEE International Conference on Acoustics, Speech and Signal Processing (ICASSP)},
  pages={12031--12035},
  year={2024},
  organization={IEEE}
}

@article{EmphaClass,
  title={Emphassess: a prosodic benchmark on assessing emphasis transfer in speech-to-speech models},
  author={de Seyssel, Maureen and D'Avirro, Antony and Williams, Adina and others},
  journal={arXiv preprint arXiv:2312.14069},
  year={2023}
}

@article{text,
  title={What do self-supervised speech models know about words?},
  author={Pasad, Ankita and Chien, Chung-Ming and Settle, Shane and others},
  journal={Transactions of the Association for Computational Linguistics},
  volume={12},
  pages={372--391},
  year={2024},
  publisher={MIT Press One Broadway, 12th Floor, Cambridge, Massachusetts 02142, USA~…}
}

@article{busso2008iemocap,
  title={IEMOCAP: Interactive emotional dyadic motion capture database},
  author={Busso, Carlos and Bulut, Murtaza and Lee, Chi-Chun and others},
  journal={Language resources and evaluation},
  volume={42},
  number={4},
  pages={335--359},
  year={2008},
  publisher={Springer}
}

@article{busso2025msp,
  title={The msp-podcast corpus},
  author={Busso, Carlos and Lotfian, Reza and Sridhar, Kusha and others},
  journal={arXiv preprint arXiv:2509.09791},
  year={2025}
}

@article{seshadri2021emphasis,
  title={Emphasis control for parallel neural TTS},
  author={Seshadri, Shreyas and Raitio, Tuomo and Castellani, Dan and others},
  journal={arXiv preprint arXiv:2110.03012},
  year={2021}
}

@article{lawrence1989concordance,
  title={A concordance correlation coefficient to evaluate reproducibility},
  author={Lawrence, I and Lin, Kuei},
  journal={Biometrics},
  pages={255--268},
  year={1989},
  publisher={JSTOR}
}

\end{document}